\documentclass[twocolumn,numberedappendix]{openjournal}

\usepackage{xcolor}
\definecolor{darkred}{rgb}{0.5, 0.0, 0.0}
\usepackage{graphicx}
\usepackage{grffile}

\usepackage{alphabeta}
\usepackage[utf8]{inputenc}
\usepackage[english]{babel}

\usepackage{hyperref}
\hypersetup{
    unicode, 
    colorlinks=true,
    linkcolor=linkcolor,
    citecolor=linkcolor,
    filecolor=linkcolor,
    urlcolor=linkcolor,
}
\usepackage{color,colortbl}
\definecolor{linkcolor}{rgb}{0.0,0.3,0.5}
\usepackage{tensind}
\tensordelimiter{?}
\DeclareGraphicsExtensions{.bmp,.png,.jpg,.pdf}
\usepackage{verbatim}
\usepackage[normalem]{ulem}
\usepackage{orcidlink}
\usepackage{soul}

\usepackage{float}
\usepackage{algorithm}
\usepackage{algpseudocode}
\usepackage{amsmath}

\newcommand{\graderr}{E_\partial}

\DeclareRobustCommand{\mfhodel}[1]{{\textcolor{darkred}{\sout{#1}}}}
\DeclareRobustCommand{\mfhorep}[2]{\ensuremath{\text{\mfhodel{$#1$}}\,\mfhoadd{#2}}}
\newcommand{\UM}{Department of Physics, University of Michigan, 450 Church St, Ann Arbor, MI 48109, USA}
\newcommand{\LCTP}{Leinweber Institute for Theoretical Physics, 450 Church St, Ann Arbor, MI 48109, USA}

\graphicspath{ {./figs/} }

\begin{document}
\title{Symbolic Regression Diagnostics of 1D Lyman-$\alpha$ Forest Emulator}

\author{Aidan Behmer\orcidlink{}}
\email{abehmer@umich.edu}
\affiliation{\UM}

\author{Ming-Feng Ho\orcidlink{0000-0002-4457-890X}}
\affiliation{\UM}
\affiliation{\LCTP}

\author{Camille Avestruz\orcidlink{0000-0001-8868-0810}}
\affiliation{\UM}
\affiliation{\LCTP}

\author{Yanhui Yang\orcidlink{0000-0001-6221-6024}}
\affiliation{Department of Physics \& Astronomy, University of California, Riverside, CA 92521, USA}

\author{Simeon Bird\orcidlink{0000-0001-5803-5490}}
\affiliation{Department of Physics \& Astronomy, University of California, Riverside, CA 92521, USA}

\begin{abstract}
We use symbolic regression to diagnose the parameter dependence of a black-box multi-fidelity emulator of the Lyman-$\alpha$ forest 1D flux power spectrum. Symbolic regression is appealing in its ability to reduce complex parameter spaces, thereby reducing compute time and offering potential interpretability in parameter inter-dependencies. Starting from a pretrained Gaussian-process (GP) emulator that connects low- and high-resolution PRIYA simulations, we search for explicit equations describing the dependence of the power spectrum on resolution and each cosmological and astrophysical parameter. We find that symbolic regression can recover equations that reproduce the emulator values well, but accurate predictions do not guarantee accurate parameter derivatives, making such equations unreliable for downstream applications such as Fisher forecasts. We therefore introduce a derivative-faithfulness diagnostic that directly compares the parameter gradient of each symbolic equation to the GP, and explore a derivative-matching (Sobolev) loss to improve this agreement. At $z=3.6$, the Sobolev objective yields derivative-faithful equations for 9 of the 11 PRIYA parameters constraining the Lyman-$\alpha$ forest, with the approach only ineffective on $h$ and $\epsilon_{AGN}$, whose 1D power spectrum imprints are weak or degenerate. 
Redefining the training objective with the Sobolev loss is a more robust and computationally efficient approach to gradient-matching repair than increasing model complexity.
Our methods present potential mitigation strategies for failure modes of PySR, enabling downstream Fisher forecasts in some of these cases.  Further, we document limitations potentially relevant to future work bridging symbolic regression with emulators. 
With this, our results demonstrate symbolic regression as an emulator diagnostic tool, with particular utility for low-dimensional or weakly coupled parameter dependencies.
\end{abstract}


\maketitle

\section{Introduction}
\label{sec:intro}

Cosmological surveys have successfully enabled precision measurements of fundamental physics using large-scale structure \citep{Sailer_2021}.  The future of cosmology, such as the further understanding of dark energy and dark matter, relies on understanding non-linear structure formation at small scales, some examples of which, specifically in Lyman-$\alpha$ structure formation, are \cite{Rogers_2021}, and \cite{Ma_2025}. One of the classic small-scale probes of cosmological structure \citep{Croft_1998} is the Lyman-$\alpha$ absorption in quasars tracing neutral hydrogen distributed around the Universe \citep{Hernquist}. 
The Lyman-$\alpha$ forest traces low-density gas, probing complementary scales of gravitational collapse to those probed in galaxy surveys. 
It is also a unique window into structure formation at redshifts $z = 2-5$, reaching scales as small as $\sim 1\,\mathrm{Mpc}$.

However, as both of those advantages highlight, the utility of the Lyman-$\alpha$ forest stems from being able to model this with a cosmological simulation, such as in \cite{Bird_2019}. For example, resolving gas on any scales smaller than 10 $\mathrm{Mpc}$ needs N-body hydrodynamic simulations as these scales are not accurately modeled by cosmological perturbation theory. To fully leverage the Lyman-$\alpha$ forest, one needs a suite of expensive simulations that span a wide range of cosmological and thermal parameters. Additionally, the Lyman-$\alpha$ forest is sensitive to smaller scales which require a higher resolution of simulations at higher redshifts (\cite{2014MNRAS.438.2499B}). Emulators are tools that make uniformly accurate estimates over a broad range of parameters by the use of simulations \citep{Bird_2019}. As a result, multi-fidelity emulation is particularly useful for Lyman-$\alpha$ studies. 

Numerous Lyman-$\alpha$ simulation efforts aim to address the need of a wide parameter range and often address this problem with using emulator-based approaches. The Lyssa suite \citep{walther} has 18 high-resolution simulations, and has a high resolution at mean density. This is a single fidelity emulator, taking a different approach to modeling larger parameter sets. Several neural network based emulators utilize Bayesian inference for their models or in the downstream applications of the model. For example, \cite{Jin_2024} uses multilayer perceptrons to build a neural network model of the relationship between a variety of thermal parameters to constrain the cosmic reionization.  \cite{Cabayol_Garcia_2023} uses a mixture density network, a type of NN that can model complex probability distributions, to model the P1D as a function of the slope and amplitude of the linear matter power spectrum rather than as a holistic function of cosmological parameters. Both \cite{Palanque_Delabrouille_2013, Palanque_Delabrouille_2015} approximate changes in the power spectrum using a Taylor expansion about the best-fit values of the cosmological parameters.

In comparison to the emulators previously described, the PRIYA \citep{2023simsuite} emulator covers parameter space with an overall higher sample density across model parameter space. PRIYA is unique in providing large-volume Lyman-alpha forest simulations with full galaxy formation physics and a realistic population of damped Lyman-alpha absorbers.
PRIYA also has already been applied to analyze eBOSS and KODIAQ-SQUAD data \citep{Fernandez.2024JCAP...07..029F,Ho.2025arXiv250918271H}.

The main advantage of a multi-fidelity emulator is to decrease computational cost by incorporating simulations with varying resolutions. Many lower resolution simulations across a wide model parameter space are combined with fewer high resolution simulations that sample a smaller subset of model parameters (\cite{2022MNRAS.509.2551H}).
With multi-fidelity methods, one can probe a wider parameter volume of the Lyman-$\alpha$ forest without having to perform a substantial amount of large simulations which are computationally expensive.
For example, a single high-resolution PRIYA simulation requires $\sim 90{\,}000$ node-hours on a supercomputer, whereas a low-resolution simulation requires $\sim 7{\,}000$ node-hours (\cite{2023simsuite}). The current best cosmological multi-fidelity emulator (\cite{Yang_2026}) can reduce this cost by up to 94\%.  Overall, multi-fidelity methods are highly effective at exploiting the resolution-volume trade-off in existing simulation suites, where different particle resolution and box sizes are used to assess numerical convergence \citep{Ho_2023}. 

However, there is a broader limitation of emulator-based inference: once constraints are derived from a given emulator, these constraints are often tied to that particular black-box model.  This coupling makes it difficult to debug and understand what has actually been learned and also limits portability. If another group uses a different emulator, it is not straightforward to test an additional parameter direction unless that degree of freedom is already built into their own simulation suite. 

One possible way to ease this limitation is to use symbolic regression to extract approximate analytical response relations from emulator outputs. These expressions would not replace the full emulator, but could provide interpretable parameter-sensitivity templates that are easier to transfer across different emulator frameworks. \cite{Kammerer_2025} shows how parameter level equations can allow for a more complete interpretation of an emulator through the individual relations.

Symbolic regression is a method of filling the data space with modeling expressions and simultaneously minimizing prediction error and model complexity (\cite{cranmer2023interpretablemachinelearningscience}).  
Symbolic regression has been applied to modeling the hypergeometric functions for the $\Lambda CDM$ comoving distance and linear growth factor (\cite{Bartlett_2026}). \cite{Bartlett_2026} also showed that one could integrate symbolic methods into a Dark Energy Survey like analysis while maintaining the accuracy of the numerical methods such as \texttt{cosmoSIS} \citep{Zuntz_2015}, \texttt{jax-cosmo} \citep{Campagne_2023}, and \texttt{CosmoPower-jax} \citep{Piras_2023}.  Similarly, \cite{Bartlett_2024} shows how a symbolic regression can be applied to the linear matter power spectrum with sub 1$\%$ accuracy. These analytic fits were able to achieve similar accuracy to black-box emulator methods in less time.  In this work, we utilize PySR\footnote{\url{https://pysr.ai/}}, a specific symbolic regression package for Python \cite{cranmer2023interpretablemachinelearningscience}. PySR has been used in numerous applications across astrophysics, such as to model galaxy-halo connections (\cite{Delgado_2022}) or to develop parametric relationships for black holes (\cite{Davis_2023}). In our application, we apply symbolic regression to the multi-fidelity emulator of the PRIYA simulation suite.

The paper is organized as follows. Section~\ref{sec:methods} describes the training data and our symbolic regression methodology. Section~\ref{sec:eqs} presents the recovered equations and the tests we put them through. Section~\ref{sec:results} reports how accurate they are, and how faithful their parameter derivatives are. Section~\ref{sec:discussion} covers the limitations of the method and its computational cost, and we conclude in Section~\ref{sec:conclusion}. Three appendices give the production PySR configuration (Appendix~\ref{sec:apdx_kwargs}), the full per-parameter equation table (Appendix~\ref{sec:apdx_per1d_eqs}), and the derivative-faithfulness approach (Appendix~\ref{sec:apdx_sobolev}).

\section{Methods}
\label{sec:methods}

In this section we first describe the PRIYA simulation suite and the 1D flux power spectrum used for training (Section \ref{sec:data}), and then introduce our symbolic regression methodology (Section \ref{sec:symbolic}) together with the Sobol-sequence sampling we use to test the resulting equations on unseen multi-dimensional combinations of parameters (Section \ref{sec:methods:sobol}).

\subsection{Data}
\label{sec:data}
The PRIYA suite is a set of cosmological hydrodynamic simulations at two resolutions run using the MP-Gadget code\footnote{\url{https://github.com/MP-Gadget/MP-Gadget}} and the galaxy formation model from Astrid \citep{2022MNRAS.512.3703B}. The lower fidelity suite  has 60 simulations with $1536^3$ gas particles, $1536^3$ cold dark matter particles and models a 120 Mpc/h periodic box with a mean interparticle spacing of 78 kpc/h. The high fidelity suite has 3 higher resolution simulations containing $2 \times 3072^3$ particles and a mean interparticle spacing of 39 kpc/$h$. The simulations are initialized at $z=99$ and run until $z=2$. The parameter space for the suite consists of 9 total simulated parameters, describing both cosmology and astrophysics, and two parameters $(\tau_0, d\tau_0)$ varied in processing to control the optical depth. 

The simulations are sampled in parameter space using the Latin Hypercube design described in \cite{2023simsuite}. There were 48 initial low-fidelity simulations, but an additional suite of 12 simulations was run in \cite{Fernandez.2024JCAP...07..029F} to better cover the posterior constraints of eBOSS DR14.

The 1D flux power spectra used to train our symbolic regression models are taken from ~\cite{Ho.2025arXiv250918271H}, which extends the PRIYA emulator to higher wavenumbers ($k = 0.003 - 0.065\,\mathrm{s/km}$) to match the smaller scales probed by the high-resolution KODIAQ-SQUAD spectra \citep{Kara_ayl__2021}. The emulator achieves a moderate precision of $\sim 5\%$--$10\%$ across the parameter space, which is sufficient for the preliminary symbolic regression study presented here.

\subsubsection{Summary of Parameters}
To provide a more complete discussion of the data, it is additionally useful to describe the parameters specific to the PRIYA suite. There are 11 parameters in total, consisting of both astrophysical and cosmological parameters. 

Following Ref.~\cite{Ho.2025arXiv250918271H}, we assume uniform priors over the parameter ranges shown in Table~\ref{tab:param_table} (for $A_P$, it is uniform (linear) prior over $A_P$):
\begin{enumerate}
    \item $d\tau_0$ :  Mean optical depth redshift evolution. 
    \item $\tau_0$ :  Mean optical depth at z = 3 relative to mean flux measurement in \cite{2007MNRAS.382.1657K}. Detailed implementation can be found in \cite{2023simsuite}.
    \item ${n_P}$ :  The Scalar spectral index. This is the slope of the primordial power spectrum at $k=0.78\, \mathrm{Mpc}^{-1}$
    \item $A_P$ :  The power spectrum amplitude at $k=0.78\, \mathrm{Mpc}^{-1}$.
    \item $z_{Hei}$ :  The start redshift of He II re-ionization.
    \item $z_{Hef}$ :  The end redshift of He II re-ionization.
    \item $\alpha_q$ :  The quasar spectral index during He II re-ionization.
    \item $h$ :  The Hubble Parameter. As discussed in \cite{2023simsuite}, $h$ only has a $1\%$ effect on the P1D for the range we consider.
    \item ${\Omega_M h^2}$ :  Total matter density. This represents all types of matter present in the data.
    \item $z_{Hi}$ :  The median redshift of the H I re-ionization
    \item $\epsilon_{AGN}$ :  The blackhole feedback factor. This is a factor varying the coupling of the thermal AGN feedback to the gas. In the ASTRID model this parameter has the value 0.05. 
\end{enumerate}
The values of these parameters are described at length in \cite{2023simsuite}. The above parameters and their maximum and minimum values are displayed in Table~\ref{tab:param_table}.

\begin{table}[!h]
\caption{\label{tab:param_table} The parameters being varied along with the ranges covered by the emulator. The Fiducial column gives the point $\theta_{\mathrm{fid}}$ at which the other parameters are held.}
\begin{ruledtabular}
\begin{tabular}{c|ccc}
Parameter & Minimum & Maximum & {Fiducial} \\
\hline
$d\tau_0$ & -0.4 & 0.25 & {$-0.009$} \\
$\tau_0$ & 0.75 & 1.25 & {1.09} \\
${n_P}$ & 0.8 & 1.05 & {0.983} \\
$A_P$ & 1.2 $\times 10^{-9}$ & 2.6 $\times 10^{-9}$ & {1.46 $\times 10^{-9}$} \\
$z_{Hei}$ & 3.5 & 4.5 & {4.0} \\
$z_{Hef}$ & 2.2 & 3.2 & {2.765} \\
$\alpha_q$ & 1.3 & 3.0 & {1.74} \\
$h$ & 0.65 & 0.75 & {0.688} \\
${\Omega_M h^2}$ & 0.14 & 0.146 & {0.1439} \\
$z_{Hi}$ & 6.5 & 8 & {7.24} \\
$\epsilon_{AGN}$ & 0.03 & 0.07 & {0.05} \\
\end{tabular}
\end{ruledtabular}
\end{table}

\subsubsection{1D Power Spectrum}
The 1D flux power spectrum (P1D) used in this work is the same as the P1D emulator used in Ref.~\cite{Ho.2025arXiv250918271H}, computed as the 1D Fourier transform along simulated sightlines of the flux contrast $\delta_F = F/\bar{F} - 1$ using \texttt{fakespectra}\footnote{\url{https://github.com/sbird/fake_spectra}}, where $F = \exp(-\tau)$ is the transmitted flux and $\bar{F}$ is the mean flux at each redshift. The mean flux (or mean optical depth) is parameterized as ratio variations from Ref.~\cite{2007MNRAS.382.1657K}, with a pivot at $z = 3$ controlled by $\tau_0$ and the redshift evolution controlled by $d\tau_0$:
\begin{equation}
\tau^{\mathrm{eff}}_{\mathrm{HI}}(z) = \tau_0 \left(\frac{1+z}{4}\right)^{3.65 + d\tau_0} \tau^{\mathrm{Kim}}_{\mathrm{HI}}(z)\,,
\end{equation}
where $\tau^{\mathrm{Kim}}_{\mathrm{HI}}(z) = 0.0023 \times (1+z)^{3.65}$, and the optical depths in the simulated spectra are rescaled in post-processing to match the desired $\bar{F}$. For the primordial power spectrum, we adopt the pivot scale $k = 0.78\,\mathrm{Mpc}^{-1}$ used in Ref.~\cite{2023simsuite}, chosen to match the scales probed by the Lyman-$\alpha$ forest and to reduce the degeneracy between $A_P$ and $n_P$.

\subsubsection{The PRIYA KODIAQ-SQUAD Emulator}
\label{sec:priya_emulator}
The training data for our symbolic regression are generated by the multi-fidelity Gaussian process emulator of \cite{Ho.2025arXiv250918271H}. The emulator extends the framework of \cite{2023simsuite} to the smaller scales needed for high-resolution P1D measurements. At each redshift, it interpolates between $60$ low-fidelity and $3$ high-fidelity PRIYA simulations using a linear multi-fidelity model, $P_F^{\mathrm{HF}}(k, \theta) = \rho \cdot P_F^{\mathrm{LF}}(k, \theta) + \delta(k, \theta)$, where $\delta$ is a per-redshift Gaussian process and $\rho$ is a per-redshift learnable parameter.

Validation uses leave-one-out cross-validation against the high-fidelity simulations \cite{2023simsuite, Ho.2025arXiv250918271H}. The median interpolation error is $\sim 1\%$ at $k = 0.01-0.06\,\mathrm{s/km}$, rising to $\sim 2\%$ at $z = 4-4.2$. Resolution convergence is checked against an extreme-resolution simulation ($2\times 768^3$ particles in a $15\,\mathrm{Mpc}/h$ box). The convergence error is $\sim 1\%$ at $k < 0.05\,\mathrm{s/km}$ and $\sim 3\%$ at $k = 0.05-0.06\,\mathrm{s/km}$. Both errors sit well below the $\sim 5\%$--$10\%$ statistical uncertainty of the KODIAQ-SQUAD measurement \cite{Ho.2025arXiv250918271H}.
Since this paper aims to use symbolic regression to extract information from the emulator and compare against the emulator itself, the accuracy of the original emulator should not affect the interpretation of the results too much.
However, the moderate accuracy at $k = 0.05-0.06\,\mathrm{s/km}$ might introduce more noise into the symbolic regression fitting, so it is expected that the symbolic regression will struggle at high $k$. The fits reported below are nevertheless restricted to $k = 0.001-0.04\,\mathrm{s/km}$ over $48$ bins, so this least accurate regime lies outside the range we actually fit.

\subsection{Symbolic Regression}
\label{sec:symbolic}
Our main tool for describing the relationship between the parameters and P1D with respect to the fidelity was PySR. PySR is a Python package that uses Symbolic Regression on data sets with a number of binary and unary operations. These operations are customizable by the user in that we are able to choose which operations the model tries to fit to the data.
This choice matters when the discovered equations are used for downstream cosmological inference. Operations that produce numerical instabilities or divergent gradients (such as unbounded division or unrestricted exponentials) can make the resulting equations unreliable for downstream analysis (e.g., Fisher forecast), even when they fit the training data well.

In our application to the PRIYA suite, we apply the model separately to every parameter within the data set.
Treating each parameter this way allows us to stress-test how reliable PySR is at extracting equations from the emulator before we trust it for a multi-dimensional fit.
In addition, PySR works best on low-dimensional problems as discussed in \cite{cranmer2023interpretablemachinelearningscience}. Conversely, we used both high- and low-fidelity parameter values simultaneously in the model, but included a resolution variable $r$ in the fitting taking the value $r=0.4$ for the LF simulations and $r=0.8$ for the HF ones.. This allows us to use both fidelities simultaneously and provides a clearer interpretation for the multi-fidelity emulator.

In what follows, we describe how the per-parameter equations are combined into a multi-parameter prediction.

\subsubsection{Normalization}
\label{sec:normalization}
For each per-parameter PySR fit we vary a single parameter $\theta_i$ across its prior while the others stay at the fiducial point $\theta_{\mathrm{fid}}$, a reference point with each parameter set to a central value within its prior range; the specific values used in this work are in our companion code repository\footnote{\url{https://github.com/jibanCat/lya1d_priya_forecast/blob/v1.2-paper/src/priya_forecast/parameters.py}, at tag \texttt{v1.2-paper} (commit \texttt{ce8ee6a}).} and include $\tau_0 = 1.09$, ${n_P} = 0.983$, $A_P = 1.46 \times 10^{-9}$, $h = 0.688$, and the corresponding central values for the IGM thermal parameters.
The $\theta_{\mathrm{fid}}$ is chosen to match \cite{Fernandez.2024JCAP...07..029F}.
We fit the \emph{logarithm} of the flux power and normalize it per $k$-bin over that parameter's sweep.
Empirically we found using the log of the flux power is easier to fit on the $A_P$ dimension.
The training target at fixed redshift $z$, wavenumber $k$, and resolution $r$ is
\begin{equation}
y(\theta_i, k, r, z) = \frac{\log P_F(\theta_i, k, r, z) - \mu_z(k)}{\sigma_z(k)}\,,
\label{eq:norm}
\end{equation}
where $\mu_z(k) = \langle \log P_F^{\mathrm{LF}} \rangle$ and $\sigma_z(k)$ are the mean and standard deviation of the low-fidelity $\log P_F$ across the parameter's prior sweep at fixed $(k, z)$. The same $(\mu_z, \sigma_z)$ are applied at both resolutions so that the resolution dependence is preserved in the residual $y$ rather than absorbed into the normalization.

The per-redshift rescaling by $\sigma_z(k)$ balances the magnitude of the residuals across the redshift range $z = 2.6$--$4.2$, since the Lyman-$\alpha$ P1D amplitude varies with the mean-flux evolution across redshift.
For the $k$ and $\theta$, we use min-max normalization: subtract the minimum and divide by the range. The purpose is to rebalance the PySR training loss across the input dimensions, so that the genetic search is not dominated by the largest-magnitude input variations.

\subsubsection{Pseudocode Model}
\label{sec:pseudocode}
The following pseudocode outlines our process of training the model on single parameter data and applying the resulting equations onto the multi-parameter data set as described in Sec.~\ref{sec:methods:sobol}. The algorithm goes as Algorithm~\ref{alg:per1d}.

\begin{algorithm}[H]
\caption{Per-parameter multi-fidelity PySR fit}
\label{alg:per1d}
\begin{algorithmic}[1]
\State \textbf{Input:} parameter set $\theta = (\theta_1, \dots, \theta_N)$, fixed redshift $z$, resolution levels $\{r_{\mathrm{LF}}, r_{\mathrm{HF}}\}$, fiducial point $\theta_{\mathrm{fid}}$
    \For {$i = 1, \dots, N$}
        \State Sample $\underline{\theta}_i$ uniformly over the $\theta_i$ prior; fix $\{\theta\}_{j \neq i} = \theta_{\mathrm{fid}}$
        \State $\underline{X} \gets \{(\underline{\theta}_i, k, r) : k \in k\text{-grid},\ r \in \{r_{\mathrm{LF}}, r_{\mathrm{HF}}\}\}$
        \State $\underline{Y} \gets \{P_F(\underline{\theta}_i, k, r)\}$ 
        \State Apply normalization (Eq.~\ref{eq:norm})
        \Repeat \Comment{up to $5$ attempts}
            \State $\mathrm{Pareto}_i \gets \texttt{PySRRegressor}(\underline{X}, \underline{Y})$ with a fresh seed
            \State $\mathcal{S}_i \gets$ Fisher-safe candidates of $\mathrm{Pareto}_i$ (Section~\ref{sec:hyperparameters})
        \Until{$\mathcal{S}_i \neq \emptyset$} \Comment{Pareto rescue}
        \State $f_i \gets \arg\min_{f \in \mathrm{Pareto}_i} \mathrm{loss}(f)$ \Comment{lowest-loss selection}
    \EndFor
    \State $f(\theta, k, r) \gets \mathrm{Combine}(\{f_i\}^N_{i=1})$ via Eq.~\ref{eq:combine}
    \State Validate: $\underline{\theta}_{\mathrm{new}} \sim \mathrm{Sobol}_N(\theta)$
    \State $\mathrm{Error} = \mathrm{MSE}\bigl(P_F(\underline{\theta}_{\mathrm{new}}, k, r),\, f(\underline{\theta}_{\mathrm{new}}, k, r)\bigr)$
\end{algorithmic}
\end{algorithm}

The production PySR kwargs used by Algorithm~\ref{alg:per1d} (operators, $\mathrm{niter}$, multithreading) are listed in App.~\ref{sec:apdx_kwargs}.

The first line declares the inputs: the parameter set $\theta$, a fixed redshift $z$, the two resolution levels $r_{\mathrm{LF}}$ and $r_{\mathrm{HF}}$, and the fiducial point $\theta_{\mathrm{fid}}$. The for-loop trains one PySR equation per parameter. For each $\theta_i$, we sample its values uniformly across its prior and keep all other parameters fixed at $\theta_{\mathrm{fid}}$. The training inputs span the full $(k, r)$ grid from the multi-fidelity emulator at the fixed redshift $z$, so each fit captures the joint dependence on the parameter, wavenumber, and resolution at once. The \texttt{PySRRegressor} call returns a Pareto front of candidate equations.
The repeat-until block is the Pareto rescue of Section~\ref{sec:hyperparameters}, and it is a test on the front rather than a rule for choosing within it. From an accepted front the selected equation $f_i(\theta_i, k, r)$ is the lowest-loss candidate; the faithfulness results of Section~\ref{sec:fisher_results} instead score the Pareto-knee candidate of the same front.

We apply this procedure to all $N=11$ parameters to obtain a set of single-parameter, multi-fidelity equations at the redshift $z$. The whole procedure is then run independently at each production redshift $z \in \{2.6, 3.6, 4.2\}$, yielding one equation set per redshift.

Outside of the for-loop, we use Sobol-sampled multi-parameter combinations to validate the learned single-parameter equations beyond their training slices. The ``Combine'' step builds a multi-parameter prediction from the per-parameter equations $f_i$, at the same fixed redshift $z$. Each $f_i$ already carries the full $(k, r)$ dependence inside it, so a naive sum would count the shared \((k,r)\) baseline multiple times. We avoid this by working with deviations from the fiducial point, in the same $\log P_F$ space in which the equations are trained (Eq.~\ref{eq:norm}).

For each parameter $\theta_i$, we define the log deviation
\begin{equation}
\Delta_i(\theta_i, k, r) = \log f_i(\theta_i, k, r) - \log f_i(\theta_{\mathrm{fid}}, k, r)\,,
\label{eq:combine_1}
\end{equation}
which depends on $\theta_i$ but vanishes when $\theta_i = \theta_{\mathrm{fid}}$. We then restore the shared $(k, r)$ baseline once, using the GP prediction at the fiducial point,
\begin{equation}
\log \bar{f}_{\mathrm{fid}}(k, r) \equiv \log P_F^{\mathrm{GP}}(\theta_{\mathrm{fid}}, k, r)\,.
\label{eq:combine_2}
\end{equation}
The combined multi-parameter prediction is
\begin{equation}
\log f(\theta, k, r) = \log \bar{f}_{\mathrm{fid}}(k, r) + \sum_{i=1}^{N} \Delta_i(\theta_i, k, r)\,,
\label{eq:combine}
\end{equation}
and the linear flux power is recovered by exponentiating, $f(\theta, k, r) = \exp[\log f(\theta, k, r)]$.

Each PySR predicted $\Delta_i$ in Eq.~\ref{eq:combine_1} is designed to vanish at $\theta_i = \theta_{\mathrm{fid}}$, so all the $(k, r)$ structure of $f_i$ that does not depend on the parameter cancels; the shared baseline is restored exactly once via Eq.~\ref{eq:combine_2}. At $\theta = \theta_{\mathrm{fid}}$ every deviation vanishes and the combine procedure returns $P_F^{\mathrm{GP}}(\theta_{\mathrm{fid}})$ exactly. A naive direct sum of the $f_i$ would have counted that shared baseline $N$ times, which is undesired.

This combine procedure is a Taylor-style expansion of $f(\theta, k, r)$ around the fiducial point in the parameter directions. Each $\Delta_i$ retains the full nonlinear dependence of $f_i$ on $\theta_i$ around $\theta_{\mathrm{fid}}$, while the cross-coupling terms $f_{ij}(\theta_i, \theta_j, k, r)$ are not modelled by construction. The limitation of this approach is discussed in Section~\ref{sec:multid_failure}.

In principle, this design aims to facilitate the use of the PySR equations in a Fisher forecast surrounding a known fiducial point (such as the Planck cosmology).
However, as we will discuss later in this draft (Section~\ref{sec:fisher_results}), making the PySR derivative faithful to a training emulator is a surprisingly challenging problem, so we do not recommend using the ``Combine'' approach beyond a pure diagnostic purpose.

\subsubsection{Hyperparameter Choices}
\label{sec:hyperparameters}
%


Our PySR hyperparameters were tuned empirically. Two of these are the number of iterations and the complexity limit we set in PySR. The number of iterations, $\texttt{niterations}$, sets how many search iterations the algorithm runs at a fixed complexity limit. The complexity limit, $\texttt{maxsize}$, is the largest complexity a candidate may reach, counted as the number of nodes in its expression tree, and limits the size of expressions PySR can search. Increasing $\texttt{niterations}$ means PySR runs more search iterations without changing the complexity limit, while increasing $\texttt{maxsize}$ allows PySR to consider larger expressions. The full production configuration is in Appendix~\ref{sec:apdx_kwargs}.

We use more iterations but a lower complexity limit than the PySR defaults: production runs $\texttt{niterations}=200$ and $\texttt{maxsize}=20$, against PySR defaults of $100$ and $30$. We deliberately choose a lower complexity limit, as each fit spans only a single 1D input parameter.
Section~\ref{sec:fisher_results} tests how far our conclusions depend on it by refitting at $\texttt{maxsize}\in\{20,30,35,40\}$. Section~\ref{sec:cost} summarizes the computational cost.


The following terminology helps clarify how we select equations in our symbolic regression pipeline:

\textbf{Pareto front}: A PySR run returns a sequence of equations with different complexities, rather than a single best equation. Simpler equations generally have higher loss, while more complex equations can give better fits. PySR keeps an equation at a given complexity only if it has a lower loss than all simpler equations, so some complexity levels may be skipped. The resulting sequence is called the \emph{Pareto front}. For example, the $\tau_0$ fit of Section~\ref{subsec:1Deqs:taup_ap} contains $18$ candidates, ranging from a one-node equation with training loss $31.7$ to a twenty-node equation with loss $0.0024$, with complexities $4$ and $12$ absent because no better equation was found there. Choosing an equation therefore means choosing how much additional complexity we are willing to trade for better accuracy. The rest of this subsection describes how we make that choice.

\textbf{Selection rules}: We also override PySR's own choice of equation from the Pareto front. The first rule is that a fit is accepted only if its Pareto front contains at least one valid equation: one that uses the parameter input, has no constant with an absolute value greater than $100$, and has a nonzero gradient with respect to the input parameter. This is deliberate because sometimes the lowest-loss equation \textit{does not use the parameter at all}.

If no valid equation is found, we re-run PySR with a new random seed. From an accepted front, we then select either the lowest-loss equation or the knee equation. The knee is the simplest valid equation whose loss is within $10\%$ of the minimum loss among valid equations.

The accuracy and prediction results use the lowest-loss equation, while we use the knee equation to report the equations in the tables and to classify each fit as \emph{faithful} or \emph{unfaithful}. We call a fit faithful if its gradient loss is below the threshold of $0.25$, and unfaithful otherwise. Appendix~\ref{sec:apdx_kwargs} states both rules and compares the knee against PySR's own default selection rule.

In this work, we have one validity filter, two ways of selecting an equation from the Pareto front, and one GP fallback rule. They are defined as follows:

\begin{itemize}
    \item \emph{Validity filter}:
    An equation is considered valid if it uses the parameter input, contains no constant with an absolute value greater than $100$, and has a nonzero gradient that can be evaluated using the finite differences in a Fisher forecast. We apply this filter when selecting the knee equation, but not when selecting the lowest-loss equation. For example, the lowest-loss $\alpha_q$ equation in Table~\ref*{tab:stats_table} contains constants $1601.11$ and $-1598.32$, and therefore does not pass this filter.

    \item \emph{The lowest-loss equation}:
    The equation with the lowest training loss on an accepted Pareto front, which is the equation stored by the PySR refit. It does not necessarily pass the validity filter.

    \item \emph{The knee equation}:
    Among the valid candidates, we select the simplest equation whose training loss is within $10\%$ of the minimum training loss among valid candidates. We use this equation to report the symbolic expressions in the tables and evaluate its gradient loss to determine whether the fit is faithful or unfaithful.

    \item \emph{GP fallback rule}:
    For the multi-parameter additive model, we also require the selected equation to pass the gradient-loss threshold of Eq.~\ref{eq:grad_err}. We test valid equations from lowest to highest loss and use the first one that passes. If none passes, we use the original GP emulator prediction for that parameter instead.
\end{itemize}

The lowest-loss and knee selections give different equations for eight of the eleven parameters at $z=3.6$. However, under the Sobolev loss, both selections give the same classification of faithful and unfaithful fits. Nine parameters pass the gradient-loss threshold: $A_P$, $\alpha_q$, $n_S$, $\tau_0$, $d\tau_0$, $\Omega_M h^2$, $z_{\rm Hei}$, $z_{\rm Hef}$, and $z_{\rm Hi}$. The remaining two, $h$ and $\epsilon_{\rm AGN}$, are unfaithful under both selections.

We run PySR in multithreaded mode using four parallel CPU cores to reduce the runtime. As a result, the genetic search is not fully deterministically reproducible between runs, even with the same random seed (see Section~\ref{sec:cost}).

\subsubsection{Operator Choices}
\label{sec:operators}

The choice of operators in the PySR search space directly affects whether the discovered equations can be used for the Fisher forecast, which requires well-defined derivatives with respect to the model parameters. We use $\{+, -, \times, \mathrm{aq}\}$ as binary operators, where $\mathrm{aq}(x,y) = x/\sqrt{1+y^{2}}$ is the analytic quotient, and $\{\exp, \log, x^2\}$ as unary operators. We list the full production configuration in Appendix~\ref{sec:apdx_kwargs}. Our earlier experimental runs included raw division $x/y$, $\sin$, $\cos$, $\sqrt{x}$, $x^{-1}$, and exponentiation $x^y$, all of which we have since dropped. We explain why we excluded these operators below, as this may also be useful for other PySR applications.

Trigonometric terms can introduce rapidly oscillating derivatives that make the resulting equations unusable for Fisher forecasts. In runs that included them, the discovered equations' parameter slopes $\partial \log P_F/\partial\theta$ differed from those of the GP by more than an order of magnitude ($\graderr\gg1$), because the fitted gradients were dominated by oscillations rather than the smooth parameter dependence of the original GP emulator.

The exponentiation operator $x^y$ allows the exponent to depend on the input variables. 
In our earlier runs, such expressions could fit the training data well but showed large deviations near the boundaries of the parameter prior (A good fit locally around the fiducial point does not necessarily mean the equation remains accurate across the full prior range). 
We dropped $\sqrt{x}$ and $x^{-1}$ for related reasons. The derivative of $\sqrt{x}$ is $1/(2\sqrt{x})$, which diverges as $x \to 0$$^+$; the derivative of $1/x$ is $-1/x^2$, which diverges as $x \to 0$.

Raw division $x/y$ can also introduce a singularity when the denominator approaches zero. Instead of removing division entirely, we replace it with the analytic quotient $\mathrm{aq}(x,y) = x/\sqrt{1+y^{2}}$. This operator behaves approximately as $x/|y|$ when $|y| \gg 1$, but its denominator never reaches zero for real inputs. This allows PySR to retain a division-like operation without introducing a pole.

The $\exp$ operator does not introduce a singularity at finite inputs. Its derivative is $\exp(x)$, which remains bounded over a finite range of $x$. Similarly, the squaring operator $x^2$ has a derivative $2x$, which is also bounded over a finite range.

\subsubsection{Loss Function}
\label{sec:loss}

PySR's default mean-squared-error (MSE) loss measures how well the predicted values match the training data, but does not directly constrain the derivatives with respect to the parameter $\theta$. When most of the variance in the training data comes from $(k, \mathrm{resolution})$, the genetic search can find equations that fit the dominant variation well without using the parameter input $\theta$ at all. We saw this happen for $6$ of the $11$ per-parameter fits at the smaller $\texttt{niterations}=20$. To address this, we re-run PySR with a new random seed if none of the equations on the Pareto front passes the validity filter described in Section~\ref{sec:hyperparameters}. This helps recover equations that depend on the input parameter.

However, including the parameter in an equation does not guarantee that its derivative is accurate. For parameters where re-running PySR is not sufficient, we instead add a derivative-matching (Sobolev) term to the training loss (Eq.~\ref{eq:sobolev}, described in Section~\ref{sec:fisher}), which penalizes the mismatch between the equation's parameter gradients and the training emulator's (GP's) gradients, in addition to the mismatch in predicted values. With this Sobolev loss, nine of the eleven parameters have faithful equations (passing 0.25 gradient loss threshold), while only $h$ and $\epsilon_{\rm AGN}$ remain above the gradient-loss threshold of $0.25$ (Section~\ref{sec:fisher}, Table~\ref{tab:faith_taxonomy}).

\section{Found Equations}
\label{sec:eqs}

\subsection{1D equations: $\tau_0$ and $A_P$}
\label{subsec:1Deqs:taup_ap}

We first illustrate the efficacy of PySR equations with the $\tau_0$ and $A_P$ variables as it is known that mean flux and primordial power spectrum amplitude are the two most important cosmological parameters in the Lyman-$\alpha$ forest 1D power spectrum.

For $\tau_0$, the PySR fit recovers a smooth, near-monotonic response in the parameter with a mild resolution coupling. In terms of the normalized inputs of Eq.~\ref{eq:norm} ($\theta$ the min-max normalized parameter, $k$ the normalized wavenumber, and $r$ the resolution feature),
\begin{multline}
f_{\tau_0}(\theta_{\tau_0}, k, r) = \bigl(\theta_{\tau_0} - e^{-1.138\,\theta_{\tau_0}}\bigr)\,\bigl(2.010 + 0.061\,r\bigr) \\
- \frac{0.511\,r}{\sqrt{k^2 + 1}} + 0.395\,.
\label{eqn:tau0_eq}
\end{multline}
The leading group $(\theta_{\tau_0} - e^{-1.138\,\theta_{\tau_0}})$ is monotonic in the parameter and scaled by a resolution-dependent term, while the second term carries a small $k$-dependent resolution offset.
For $A_P$, the amplitude enters quadratically,
\begin{equation}
f_{A_P}(\theta_{A_P}, k, r) = \theta_{A_P}^{2} - \frac{\bigl(1.276 - \theta_{A_P}\,e^{k}\bigr)^{2}\,(r + 0.959)^{2}}{\bigl(e^{k^{2}} - 1.822\bigr)^{2} + 1}\,,
\label{eqn:ap_eq}
\end{equation}
a squared $(1.276 - \theta_{A_P}\,e^{k})$ group whose $k$-envelope steepens the amplitude sensitivity toward smaller scales, modulated by resolution through the $(r + 0.959)^{2}$ factor.

For $\tau_0$, the lowest-loss and Pareto-knee selections (described in Sec~\ref{sec:hyperparameters}) coincide at complexity $20$. For $A_P$, they differ by only one node: the lowest-loss equation above has the amplitude entering quadratically, while the complexity-$19$ knee is linear at leading order. Both, nevertheless, satisfy the derivative-faithfulness criterion. 

\begin{figure*}
\centering
\includegraphics[width=0.95\linewidth]{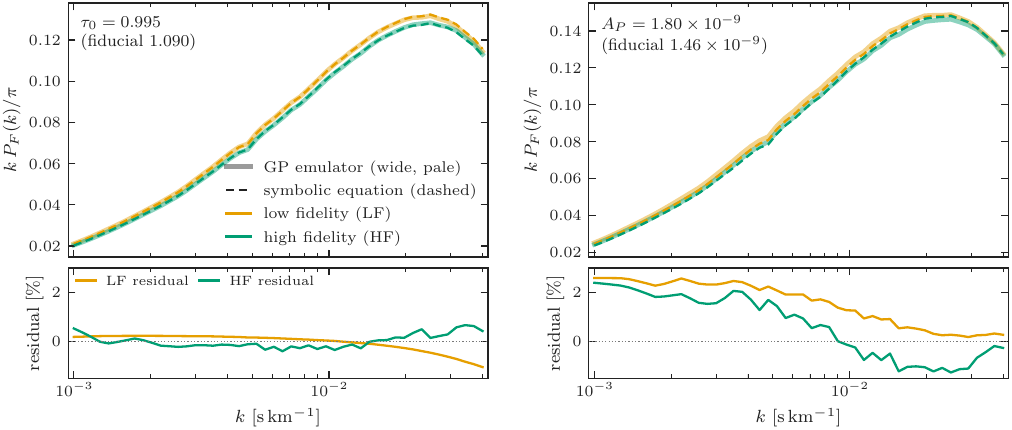}
\caption{
Comparison of the 1D PySR predictions with the original GP emulator at $z=3.6$, varying $\tau_0$ (left; Eq.~\ref{eqn:tau0_eq}) and $A_P$ (right; Eq.~\ref{eqn:ap_eq}). Top: $k\,P_F(k)/\pi$ predicted by the GP (solid lines) and PySR (dashed lines), with low-fidelity (LF) in orange and high-fidelity (HF) in green. Bottom: relative error (residual), $1-P_F^{\mathrm{PySR}}/P_F^{\mathrm{GP}}$, with LF in,orange and HF in,green. \footnote{The figure is regenerated via \texttt{scripts/regen\_fig1.py} in this manuscript's repository, from the single-redshift Sobolev refits at $z=3.6$.} The PySR predictions use the lowest-loss equations (Section~\ref{sec:hyperparameters}).}
\label{fig:tau0_ap_pred}
\end{figure*}

Figure~\ref{fig:tau0_ap_pred} compares the prediction from Eqs.~\ref{eqn:tau0_eq} and~\ref{eqn:ap_eq} against the GP emulator at both fidelity levels, plotting $\tau_0$ and $A_P$ each at a single non-fiducial value while all other parameters are kept at $\theta_{\mathrm{fid}}$.

The complete listing of all $11$ per-parameter equations, together with their complexity, training loss, and LF/HF relative errors, is given in App.~\ref{sec:apdx_per1d_eqs}.

With predictions from the GP emulator for each parameter, we computed the percentage error, as shown in Table \ref{tab:stats_table}. 
These tests indicate that ${n_P}$ was the worst parameter and $h$ was the second worst, with the percentage error values for each shown in Table \ref{tab:stats_table}. This is not a surprise because strong constraint on $n_S$ requires accurate predictions at both low-$k$ and high-$k$, so $n_S$ by construction is harder to predict by an emulator trained on simulations of finite resolution and non-zero cosmic variance. The Hubble constant, $h$, in contrast, is hard to predict by P1D emulator because the data vector lives in $\mathrm{km}/\mathrm{s}$ space while simulation resolution lives in $\mathrm{Mpc}\,h^{-1}$ space. So when $h$ vary in the input parameter, some part of the largest scales, which contains the most of the comic variance, move in and contaminate the largest mode of the emulator in $\mathrm{km}/\mathrm{s}$ unit (this is mentioned in \cite{2023simsuite})

\begin{table}[!h]
\caption{\label{tab:stats_table} Mean and maximum relative percent error of the per-parameter 1D PySR equations at $z=3.6$ on the per-parameter prior sweep used for training ($50$ points across the prior of the varied parameter, all others held at the fiducial), evaluated at low and high fidelity. Rows are sorted by descending high-fidelity mean error. Selection: the lowest-loss equation (Section~\ref{sec:hyperparameters}). Last row: the GP's own leave-one-out error against the simulations (an over-estimate for HF, which has only $3$ simulations).}
\begin{ruledtabular}
\centering
\setlength{\tabcolsep}{4pt}   
\begin{tabular}{c|cccc}
$\theta$ & \shortstack{LF mean\\(\%)} & \shortstack{LF max\\(\%)} & \shortstack{HF mean\\(\%)} & \shortstack{HF max\\(\%)} \\
\hline
${n_P}$ & 0.86 & 5.30 & 5.16 & 10.78 \\
$h$ & 0.21 & 1.72 & 1.68 & 10.50 \\
$z_{Hi}$ & 0.10 & 1.28 & 1.61 & 7.76 \\
$\alpha_q$ & 1.24 & 1.95 & 1.46 & 3.18 \\
$A_P$ & 1.64 & 4.49 & 1.18 & 4.49 \\
${\Omega_M h^2}$ & 0.17 & 2.02 & 1.16 & 5.28 \\
$z_{Hef}$ & 0.59 & 2.77 & 1.07 & 7.41 \\
$\epsilon_{AGN}$ & 0.10 & 1.20 & 0.38 & 2.09 \\
$z_{Hei}$ & 0.11 & 0.81 & 0.29 & 1.36 \\
$\tau_0$ & 0.22 & 1.04 & 0.24 & 1.21 \\
$d\tau_0$ & 0.01 & 0.02 & 0.11 & 0.54 \\
\hline
{GP (LOO)} &{0.71} & {8.91} & {0.58} & {3.69} \\
\end{tabular}
\end{ruledtabular}
\end{table}

\subsection{Sobol Sequence Validation}
\label{sec:methods:sobol}
So far, we have demonstrated performance of models trained to fit PySR to the PRIYA emulator's per-$\theta$ output. We have not yet validated the multi-dimensional performance of the ``Combine'' steps in Eq~\ref{eq:combine_1} and Eq~\ref{eq:combine_2}.
To expand our model, we make predictions from an existing PRIYA emulator by sampling the input parameters in a multi-dimensional Sobol sequence $(N=256)$. This creates a separate testing set not used in the PySR training.
The resulting data set is such that only the selected parameters are varied across the P1D spectrum and $k$ values. We used this in conjunction with the equations outputted from the PySR to model the predicted P1D against the GP's actual values.
By showing the results of the worst 2D/3D combination, we demonstrate the worst-case scenario of our simple ``Combine'' method.

\begin{figure}
\centering
\includegraphics[width=\linewidth]{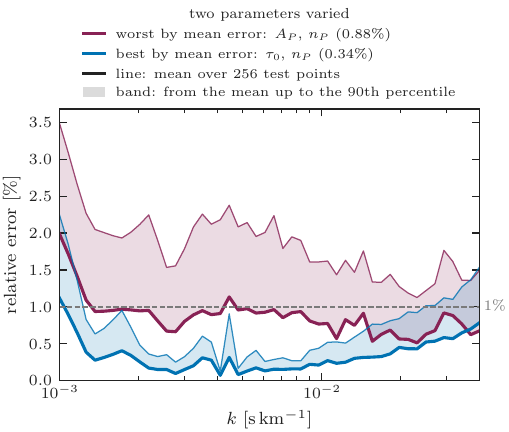}
\caption{
Comparison of the additive model against the GP at $z=3.6$, varying two parameters at a time. We show the combinations with the lowest ($\tau_0$, $n_P$) and highest ($A_P$, $n_P$) mean relative errors from Table~\ref{tab:multid}, evaluated at $256$ Sobol points for each comparison. The solid lines show the mean relative error at each $k$, while the shaded regions extend to the 90th percentile over the test points. The dashed line marks $1\%$ error. Both combinations have mean errors below $1\%$ when averaged across the test points and $k$-bins, although larger errors appear in the 90th percentile, particularly at small $k$. The corresponding three-parameter results are shown in Figure~\ref{fig:multid_3d}.
}
\label{fig:multid_bestworst}
\end{figure}

\begin{figure}
\centering
\includegraphics[width=\linewidth]{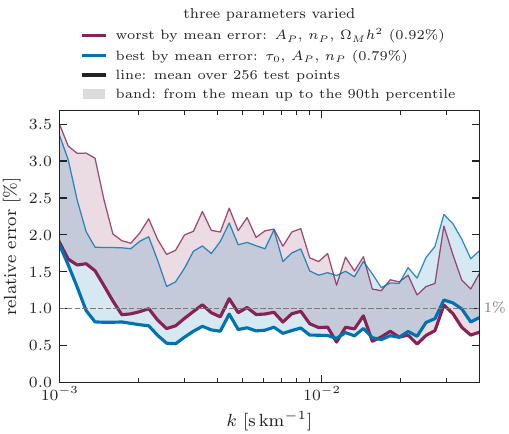}
\caption{
Same as Figure~\ref{fig:multid_bestworst}, but varying three parameters at a time. We show the combinations with the lowest and highest mean relative errors from Table~\ref{tab:multid}. The lines show the mean relative error across $256$ Sobol test points at each $k$, with shaded bands extending to the 90th percentile. The dashed line marks $1\%$ error.
}
\label{fig:multid_3d}
\end{figure}

Figure~\ref{fig:multid_bestworst}, Figure~\ref{fig:multid_3d}  and Appendix~\ref{sec:apdx_multid}'s Table~\ref{tab:multid} show the validation of the combined symbolic equations on a Sobol sequence that was not seen during training. We will discuss per-parameter accuracy results in Section~\ref{sec:well_results} in full.
We limit ourselves to plot against 2D or 3D combination.
For each parameter combo we plot the relative error of the combined symbolic prediction against the GP emulator across the Sobol points, with the $1\%$ reference line indicated.


Table~\ref{tab:multid} shows that the cosmology mean-flux block ($A_p,{n_P},\tau_0$) performs better than the IGM-thermal block ($z_{Hei},z_{Hef},\alpha_q$), which is at worst $\sim 10\%$ relative error for the 3D combo. The worst-performing 2D and 3D parameter combinations are shown in Table~\ref{tab:multid}, with the $k$ of the worst bin indicated.
This is interesting and expected because the IGM thermal model parameters are from the same recipe in \cite{Phoebe:2020MNRAS.496.4372U}, which control the initial/final redshifts of HeII reionization and its quasar spectral index, and are inherently correlated.
Our ``Combine'' method omits the cross-coupling terms between these parameters, which is why the relative error is higher than for the cosmology block.
The worst performing scale, ${k_{\rm worst}}$ is always at the low-$k$ end.
This is also expected because the finite box size of the PRIYA suite naturally limits the number of low-$k$ modes, which increases the variance of the P1D at low-$k$ and makes it more difficult for the PySR to fit the low-$k$ behavior.

\subsection{Resolution dependence}
\label{sec:resolution}

The extracted equations are naturally multi-fidelity: each per-parameter equation carries a resolution
feature $r$ that labels the fidelity of the training data. The PySR fits LF and HF data jointly and interpolates between them.
The above multi-dimensional combination is always evaluated at the HF anchor $r=0.8$ (to obey the convention of normalization in range $(0, 1)$).
We emphasize that $r$ is sampled only at these two anchors rather than swept
continuously, so it encodes a fidelity correction rather than a physical
resolution scale.
Thus, the equations are not intended for extrapolation beyond
$r\in\{0.4,0.8\}$ though it is a natural extention for future work.

Every one of the eleven production equations
(Table~\ref{tab:per1d_eqs}) retains an explicit $r$ dependence: the symbolic
search did not discard the resolution feature for any parameter.
Interestingly, most of the $r$ coupling is with $k$ in some way, only ${\Omega_M h^2}$ and $z_{\mathrm{Hi}}$ have mostly $k$-independent $r$ terms.
We characterize it through the per-parameter
multiplicative lift at the fiducial point,
\begin{equation}
    \begin{split}        
  R_i(k) &= \frac{P_F^{\mathrm{HF}}(\theta_{\mathrm{fid},i},k)}
                {P_F^{\mathrm{LF}}(\theta_{\mathrm{fid},i},k)}\\
         &= \exp\!\Bigl\{\bigl[f_i(x_0,x_1,0.8) - f_i(x_0,x_1,0.4)\bigr]\,\sigma_z(k)\Bigr\},
  \label{eq:rescorr}
    \end{split}
\end{equation}
The equations are trained on the normalized logarithm of Eq.~\ref{eq:norm}, so the ratio of powers is an exponential of a difference rather than a ratio of the de-normalized outputs. The per-$k$ offset $\mu_z(k)$ is common to both fidelities and cancels, leaving the correction set entirely by the per-$k$ scatter $\sigma_z(k)$ and by how far the equation moves between the two resolution anchors.
For the cosmology and mean-flux parameters at $z=3.6$, $R_i(k)$ shows that the ratio departs from unity by $4\%$ for $d\tau_0$ rising to $15\%$ for $h$, and depends on both $k$ and the value of the varied parameter.

\subsection{Derivative loss}
\label{sec:fisher}
We compare the symbolic emulator to the Gaussian-process (GP) emulator not only on the \textit{value} of the power spectrum but also on its \emph{slope}, the parameter derivative $\partial P_F / \partial \theta$ that a downstream emulator analysis, e.g., Fisher forecast, would consume. The reason is that an equation can reproduce $P_F$ to a percent and still bias that derivative if its slope is wrong (similar to ``Fisher's Mirage'' mentioned in \cite{Wilson:2025PhRvD.111j3532W}). We therefore score each equation's slope against the GP's,
\begin{equation}
{\graderr} = \operatorname{median}_{k}
\left| \frac{\partial_\theta \log P_F^{\mathrm{eq}}}{\partial_\theta \log P_F^{\mathrm{GP}}} - 1 \right|\,,
\label{eq:grad_err}
\end{equation}
a ratio of central finite-difference slopes of $\log P_F$ at the fiducial point (over $k$-bins where the GP slope is non-negligible), and call an equation faithful (i.e., a good representation of the training emulator) when ${\graderr} \le 0.25$.

A PySR run returns a Pareto front of candidate equations for each parameter, trading equation complexity against training loss, so we must state which candidate we score. We report the \emph{Pareto-knee} equation defined in Section~\ref{sec:hyperparameters}.

Every faithfulness number we quote, in Table~\ref{tab:faith_taxonomy} and Table~\ref{tab:per1d_eqs}, uses this Pareto-knee selection.

To \emph{cure} an unfaithful PySR equation we add a derivative-matching (Sobolev) term to the training loss,
\begin{equation}
L = \big\langle (y_{\mathrm{eq}} - y_{\mathrm{GP}})^2 \big\rangle
+ \lambda\, \big\langle (\partial_\theta y_{\mathrm{eq}} - \partial_\theta y_{\mathrm{GP}})^2 \big\rangle\,,
\qquad \lambda = 5\,,
\label{eq:sobolev}
\end{equation}
on the normalised (log-$P$) training residual $y$ of Eq.~\ref{eq:norm}. Since the residual $y$ is a rescaled $\log P_F$, the Sobolev term matches the same log-derivative that Eq.~\ref{eq:grad_err} scores.

The hyperparameter $\lambda$ is a tunable weight on the Sobolev term.
We empirically tuned it to $\lambda=5$ to recover the PySR equations we think are faithful enough to the training emulator.
The resolution here with the Sobolev term is a bit ad-hoc as we are scoring the equations against the gradient error.
However, the found equations with the Sobolev term are at least much more faithful in terms of getting the parameter constraints.
We discuss in more detail the effect of the Sobolev term in Section~\ref{sec:results}.

\section{Results}
\label{sec:results}
We test the symbolic model on variations of one, two, and three parameters, and report the errors separately for the low- and high-fidelity regimes. Table~\ref{tab:stats_table} summarizes the one-dimensional errors for each parameter, while Table~\ref{tab:multid} shows the errors after combining multiple parameter dependences through the additive-Taylor construction. Some parameter dependences are recovered cleanly, while others are noticeably harder to model (Section~\ref{sec:well_results}). We will then test whether equations that reproduce the P1D well also preserve the parameter derivatives needed for any downstream inference tasks (Section~\ref{sec:fisher_results}).

\subsection{Well-predicted \& Challenging parameter variations} 
\label{sec:well_results}

We briefly summarize the accuracy of the per-parameter one-dimensional equations and their multi-parameter combination.

\begin{figure}
\centering
\includegraphics[width=\linewidth]{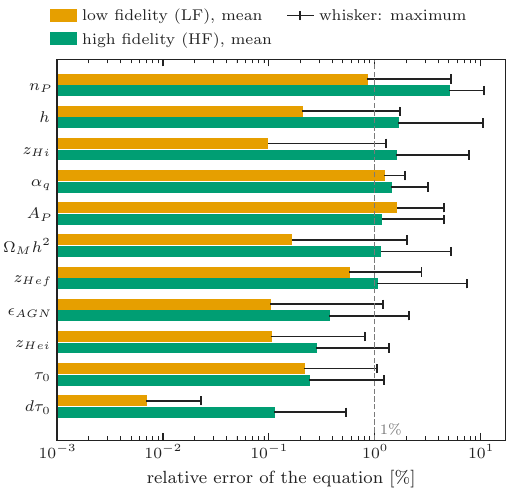}
\caption{
Mean and maximum relative errors of the 1D PySR equations at $z=3.6$, comparing low-fidelity (LF, orange) and high-fidelity (HF, green) predictions. The bars show the mean relative error across each parameter's prior range, while the whiskers indicate the maximum error. The dashed line marks $1\%$. The eleven parameters are ordered by decreasing HF mean error, following Table~\ref{tab:stats_table}. All results use the lowest-loss equation (Section~\ref{sec:hyperparameters}).}
\label{fig:param_error_bars}
\end{figure}

We first discuss how well PySR can recover the 1D parameter variations.
Figure~\ref{fig:param_error_bars} shows the mean and maximum relative error of the 1D equations at $z=3.6$, evaluated at both fidelities on the per-parameter prior used for training ($50$ points across the prior of the varied parameter, all others held at the fiducial).
In general, at LF, the mean prediction errors are roughly below $1\%$.
At HF, the worst prediction errors could reach $10\%$ for the most challenging parameters.

The best-performing parameter dimensions at HF are the mean-flux slope $d\tau_0$ and the mean-flux amplitude $\tau_0$.
This makes sense because the dependence on $\tau_0$ is almost linear in each redshift bin.

The large variation in P1D also makes it easier for the PySR to find a good fit.

In contrast, the worst performing parameters are ${n_P}$, $h$, and $z_{Hi}$.
In \cite{2023simsuite}, the PRIYA simulations show that the 1D flux power spectrum is not very sensitive to $h$ and $z_{Hi}$.
$h$ represents the change of $km/s$ to $Mpc/h$ conversion, while largely affected by the cosmic variance of the simulation box at low $k$, and thus the effect is not smooth. $z_{Hi}$ is the redshift of hydrogen reionization, which is higher redshift ($z_{Hi} > 5$) than the nominal redshift of the P1D ($z=2.2-4.6$) of either eBOSS or KODIAQ-SQUAD, and thus the P1D is not very sensitive to this parameter.
${n_P}$ is the slope of the primordial power spectrum, which is the primary cosmological parameter that P1D is constraining.
The fact that it's hard to use PySR to find a good fit for ${n_P}$ is suggesting this ${n_P}$ dimension is more complex than the other parameters. It is also possible the ${n_P}$ dimension is influenced by PySR struggling with fitting slope changes at both fidelities and therefore causing an effect at low and high k values each. 
This makes sense because P1D is as a function of $k$ scale, so intrinsically the low-$k$ end has more cosmic variance and high-$k$ end has more resolution effects, while a good ${n_P}$ fit requires a good fit across the entire $k$ range.
We will also mention the difficulty of making reasonable gradients for ${n_P}$ in Section~\ref{sec:fisher_results}.

We briefly discuss the multi-parameter combination from extrapolation of the additive-Taylor method using our 1D PySR equations.
We use the method in Algorithm~\ref{alg:per1d} to combine the 1D equations into a multi-parameter prediction, and predict against the unseen Sobol sequence of $256$ points per combination.
The goal is to show how much information we capture or lose when we only vary in per-dimension manner.
Figure~\ref{fig:multid_bestworst} (with Figure~\ref{fig:multid_3d} for three varied parameters) summarizes the additive-Taylor combination method comparison against the GP over $256$ Sobol draws per combination.
We specifically tested against the combinations of cosmology, mean flux, and IGM thermal parameters.
Most of the tested combinations are below $1\%$ mean error, which is unsurprising because the Taylor-additive method is expected to perform fine around the fiducial parameter point and the mean error is dominated by the Sobol points that are close to the fiducial point.
The useful information is in the maximum error, which shows how much information we lose when we only vary in per-dimention manner (Table~\ref{tab:multid}).
Interestingly, the He\,II reionization combinations carry the largest tails in terms of maximum error (Table~\ref{tab:multid}).
This makes sense because the additive-Taylor method drops the cross-coupling between $z_{Hei}$, $z_{Hef}$, and $\alpha_q$: the 2D $z_{Hei}$--$\alpha_q$ pair reaches $8.70\%$ and the 3D $z_{Hei}$--$z_{Hef}$--$\alpha_q$ combination $9.79\%$. 
Because the He\,II reionization parameters are from the same recipe as \cite{Phoebe:2020MNRAS.496.4372U}, they are expected to be correlated and thus the additive-Taylor method is not able to capture this correlation.

\subsection{Derivative error}
\label{sec:fisher_results}

Value accuracy, such as the one shown in Figure~\ref{fig:param_error_bars}, alone does not fully validate that the PySR equations can give reasonable parameter constraints.
For example, $\epsilon_{AGN}$ has the fourth-smallest high-fidelity mean error in Table~\ref{tab:stats_table} at $0.38\%$, but its slope/gradient error under the Sobolev objective is $0.771$, well above the $0.25$ threshold for a derivative-faithful equation.

\begin{figure*}
\centering
\includegraphics[width=\linewidth]{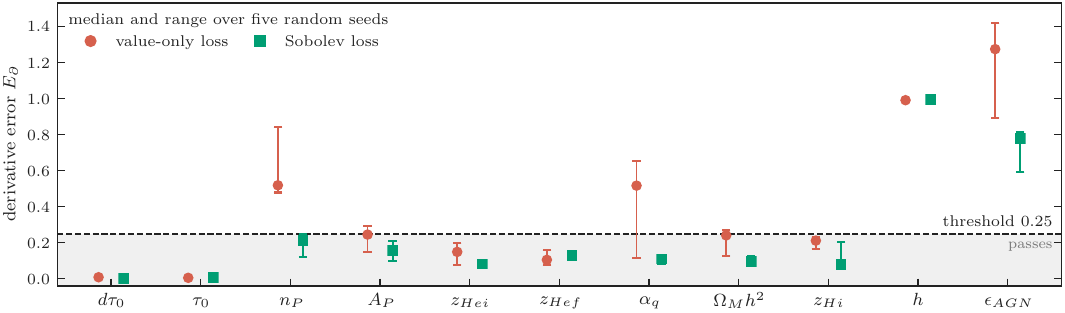}
\caption{
Derivative errors across five random seeds (0--4) at $z=3.6$, comparing value-only (circles) and Sobolev (squares) searches. Each point shows the median $\graderr$ of the Pareto-knee equations across seeds, with whiskers indicating the minimum and maximum. The dashed line marks the $0.25$ threshold for a faithful equation. With the Sobolev loss, most parameters have small derivative errors that remain below the threshold across all five seeds, including $n_P$ (median $0.21$). In contrast, the value-only fits show larger variations between seeds. The exceptions are $h$ and $\epsilon_{\mathrm{AGN}}$, which remain above the threshold for all five seeds even with the Sobolev loss. The $h$ value-loss point has only one surviving seed and is therefore shown without whiskers. All results use the Pareto-knee selection (Section~\ref{sec:hyperparameters}).
}
\label{fig:seed_band}
\end{figure*}

\begin{table*}[htbp]
\caption{
Per-parameter derivative errors at $z=3.6$. The table compares the derivative error $\graderr$ (Eq.~\ref{eq:grad_err}) of the Pareto-knee equations under the value-only and Sobolev losses. For both losses, we also report the median and $[\min,\max]$ across five random seeds (0--4). Equations with $\graderr<0.25$ are classified as faithful. All results use the Pareto-knee selection (Section~\ref{sec:hyperparameters}).
}
\label{tab:faith_taxonomy}
\begin{ruledtabular}
\begin{tabular}{l|cc|cc|c}
 & \multicolumn{2}{c|}{{single run (main fit)}} & \multicolumn{2}{c|}{{five seeds: median $[\min,\max]$}} & \\
parameter & value & Sobolev & {value} & Sobolev seed band & class \\
\hline
$d\tau_0$ & $0.009$ & $0.003$ & {$0.009\,[0.009,0.009]$} & $0.003\,[0.002,0.003]$ & faithful \\
$\tau_0$ & $0.006$ & $0.005$ & {$0.005\,[0.005,0.007]$} & $0.005\,[0.004,0.006]$ & faithful \\
${n_P}$ & $0.365$ & $0.160$ & {$0.519\,[0.479,0.841]$} & $0.212\,[0.123,0.246]$ & faithful \\
$A_P$ & $0.298$ & $0.155$ & {$0.245\,[0.147,0.295]$} & $0.158\,[0.100,0.212]$ & faithful \\
$z_{Hei}$ & $0.079$ & $0.084$ & {$0.149\,[0.078,0.198]$} & $0.081\,[0.078,0.087]$ & faithful \\
$z_{Hef}$ & $0.105$ & $0.045$ & {$0.105\,[0.075,0.159]$} & $0.130\,[0.112,0.142]$ & faithful \\
$\alpha_q$ & $0.126$ & $0.134$ & {$0.517\,[0.113,0.654]$} & $0.110\,[0.081,0.132]$ & faithful \\
${\Omega_M h^2}$ & $0.697$ & $0.063$ & {$0.241\,[0.129,0.271]$} & $0.096\,[0.087,0.126]$ & faithful \\
$z_{Hi}$ & $0.266$ & $0.167$ & {$0.212\,[0.166,0.234]$} & $0.078\,[0.077,0.204]$ & faithful \\
$h$ & $0.986$ & $1.000$ & {$0.991$ (one seed)} & $0.996\,[0.976,1.000]$ & resistant \\
$\epsilon_{AGN}$ & $1.418$ & $0.771$ & {$1.275\,[0.895,1.418]$} & $0.778\,[0.590,0.815]$ & resistant \\
\end{tabular}
\end{ruledtabular}
\end{table*}

Figure~\ref{fig:seed_band} summarizes the per-parameter faithfulness verification across five PySR randomness seeds.
A gradient error threshold, $0.25$, is designed to visually indicate whether a candidate equation is faithful enough to behave as a good approximation of the target emulator.
Interestingly, relying solely on value loss (red dots) cannot reproduce the same level of faithfulness at the gradient level,
which implies at these dimensions, although the emulator predictions are good but the gradients are not well captured enough to be used for downstream inference analysis.
One obvious plausibility is that even though one can have a good MSE value loss, it might just happen by chance that the value variation is too small to be captured by the PySR equation.
Their gradients might be arbitrarily larger or smaller, making any downstream inference analysis biased.
Figure~\ref{fig:seed_band} shows that the Sobolev loss (green squares) is able to recover the gradient error threshold for most parameters, including ${n_P}$, by including the gradient into the loss function when searching for the symbolic equations.

\begin{figure*}
\centering
\includegraphics[width=0.95\linewidth]{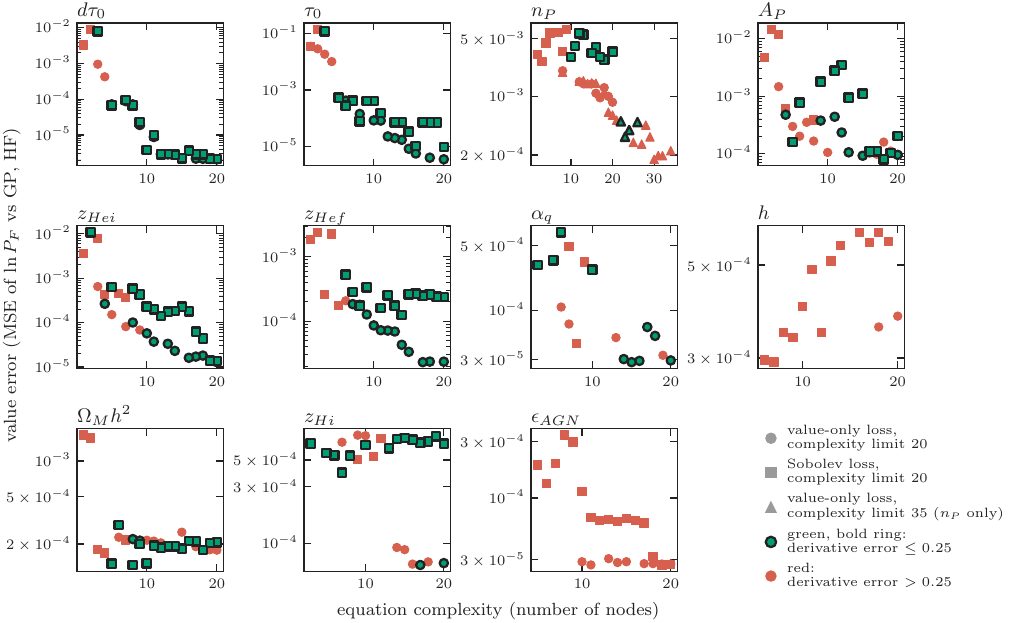}
\caption{
Per-parameter derivative errors at $z=3.6$. Each marker represents a candidate PySR equation, with its complexity on the $x$-axis and prediction error on the y-axis (log scale; lower is better). The colour indicates the derivative error $\graderr$ (green: faithful; red: unfaithful), while a bold ring highlights candidates that pass the $0.25$ threshold. Circles: value-only search; squares: Sobolev search (which includes a gradient-matching term in the loss; Eq.~\ref{eq:sobolev}). Triangles show the value-only search with a higher complexity limit for $n_P$. The figure includes only candidates that pass the validity filter (Section~\ref{sec:hyperparameters}): $331$ of the $416$ candidates across these Pareto fronts, with the remaining $85$ excluded.
}
\label{fig:pareto_faith}
\end{figure*}

To study how the search budget affects PySR equations' faithfulness towards gradient errors, we show the PySR's loss versus model complexity in Figure~\ref{fig:pareto_faith}.
In Figure~\ref{fig:pareto_faith}, one panel per parameter, every candidate equation placed by complexity and value error and \emph{coloured by its gradient error} (Eq.~\ref{eq:grad_err}), with a bold black ring on equations that clear the $0.25$ threshold.
Under the Sobolev loss function, nine of the eleven parameters are faithful to gradient error at the Pareto knee \emph{at $z=3.6$} (Table~\ref{tab:faith_taxonomy}): the mean-flux pair ($d\tau_0$, $\tau_0$), the primordial amplitude and tilt ($A_P$, ${n_P}$), the He\,II reionization block ($z_{Hei}$, $z_{Hef}$, $\alpha_q$), and the matter density and hydrogen reionization redshift ($\Omega_M h^2$, $z_{Hi}$). Five of these are already faithful on the value loss, while ${n_P}$ becomes faithful once the Sobolev slope-matching term is added to the loss function at the same $\mathrm{maxsize}=20$ budget.

The case of $A_P$ is selection-sensitive: a faithful, lower-complexity $A_P$ equation already exists using value loss, while adding the Sobolev term produces non-monotonic behaviour in the value loss versus complexity curve.
Only $h$ and $\epsilon_{\mathrm{AGN}}$ resist every method: their P1D response at the fiducial point is too weak or too degenerate to pass the gradient error threshold. The result is moderately redshift-dependent: retraining the Sobolev fits at $z=2.6/3.6/4.2$ gives $8/9/6$ faithful parameters, and only five ($d\tau_0$, $\tau_0$, $A_P$, $\Omega_M h^2$, $z_{Hi}$) are faithful at all three redshifts; ${n_P}$ is faithful only at $z=3.6$ (${\graderr}=0.255/0.160/0.379$ at $z=2.6/3.6/4.2$) and is borderline there (five-seed band maximum $0.246\approx$ threshold).

\begin{figure}
\centering
\includegraphics[width=\linewidth]{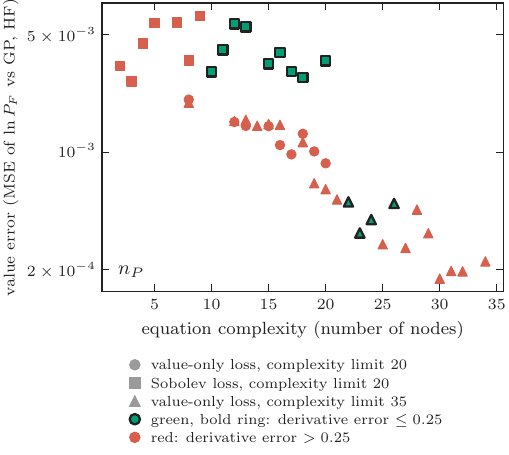}
\caption{
For $n_P$, the value-only search (triangles) reduces the prediction error but only passes the $0.25$ gradient-loss threshold at $\mathrm{maxsize}\ge35$, and only for some random seeds. In contrast, the Sobolev search (squares) already finds equations that pass the threshold at the smallest complexity limit, $\mathrm{maxsize}=20$. The colours indicate whether each candidate passes the ``threshold''. As in Figure~\ref{fig:pareto_faith}, we show the $47$ valid candidates out of the $66$ candidates on these Pareto fronts.
}
\label{fig:ns_budget}
\end{figure}

As ${n_P}$ is an important cosmological parameter, we further study its robustness to PySR randomness and search budget, which is shown in Fig.~\ref{fig:ns_budget}.
PySR is stochastic and the complexity budget is a tunable hyperparameter, so we retrained the value and Sobolev losses at five seeds and vary the complexity budget $\mathrm{maxsize}\in\{20,30,35,40\}$. Two patterns emerge. First, value-loss faithfulness is budget- and seed-sensitive: the ordinary ${n_P}$ search clears the threshold only once the budget is raised to $\mathrm{maxsize}\ge35$ (${\graderr}$ falls from $0.365$ at $\mathrm{maxsize}=20$ to $0.149$ at $35$), and even then only two of five seeds clear it (median $0.27$, range $[0.06,0.31]$).

Second, the Sobolev objective reaches the faithfulness threshold already at the smallest tested budget, $\mathrm{maxsize}=20$, and does so consistently across seeds (${n_P}$ median $0.212$, with all five seeds $\leq0.25$). In this case, adding the derivative term therefore makes the search more effective than the value-only loss at the same budget. This improvement is not universal, however: $h$ and $\epsilon_{\mathrm{AGN}}$ remain above the threshold for every seed and every budget we test.

\section{Discussion}
\label{sec:discussion}

\subsection{Limitations of joint multi-parameter fits}
\label{sec:multid_failure}
The additive-Taylor construction treats each parameter dependence separately, and therefore cannot capture explicit couplings between parameters. We tested whether these couplings could instead be learned by fitting a single symbolic equation over six parameters and $k$, at $z=3.6$ and high fidelity. This turns out to be a much harder problem. At a modest complexity budget, the joint equations reproduce the emulator while retaining only a few independent parameter directions in their gradients.

The equations themselves show how this information is lost. Some parameters are forced to share the same dependence on $k$, making their gradients nearly proportional. In other cases the parameter dependence becomes independent of $k$, or a weak parameter is dropped from the equation altogether. These different forms lead to the same basic failure: a good fit to the power spectrum does not guarantee that the joint parameter dependence has been preserved. For AGN feedback this is not necessarily as important as at $z>2$ the AGN feedback does not significantly impact the Lyman-$\alpha$ P1D \citep{tillman2025effectsagnfeedbacklymanalpha}. Similarly, $h$ has a minimal effect on the P1D for the range considered \citep{2023simsuite}. 

Increasing the complexity budget recovers most of the missing gradient directions. The Jacobian grows from only a few independent directions at $\mathrm{maxsize}=25$ to about five of six by $\mathrm{maxsize}=100$. The recovery is not complete, and the largest fits are also more difficult to search reliably. We therefore view the joint fit as a useful demonstration that symbolic regression can begin to recover coupled parameter dependence, but that doing so faithfully requires substantially more complexity than the one-dimensional fits. For the present analysis, the additive-Taylor construction remains the more controlled approach.

\subsection{Computational cost and debugging}
\label{sec:cost}
Each PySR fit performs a genetic search over equation space and is intrinsically more expensive than a simple polynomial fit. A production fit runs at $\mathrm{niterations}=200$, $\mathrm{populations}=48$ and $\mathrm{maxsize}=20$ in multithreaded mode on four cores with $8$\,GB of memory. Under the Sobolev loss (gradient-matching) a fit takes a median of $13.1$\,min ($9.3$--$21.1$\,min over the $87$ fits at that budget). As a comparison, a plain value loss (without gradient-match) at the identical budget the median is $4.2$\,min. So the derivative matching costs about a factor of three in wall time.

Cost grows with the complexity cap: raising $\mathrm{maxsize}$ from $20$ to $30$ and then to $40$ moves the mean fit from $13.6$ to $18.3$ and $23.8$\,min under the Sobolev loss, and from $4.3$ to $6.5$ and $9.1$\,min under the plain value loss. The campaign as a whole, $254$ fits, cost $177$ core-hours.

Joint fits over multi-dimensional subsets are substantially harder. A single $\mathrm{niter}=100$, $\mathrm{maxsize}=25$ fit over the six-parameter subset takes $17.1$\,min on a 16-CPU allocation ($25.2$\,min on four cores). The complexity sweep of Section~\ref{sec:multid_failure} is more expensive again: at $n_{\mathrm{train}}=512$ and $400$ iterations, with four CPUs allocated per fit, one seed runs $43$--$108$\,min at $\mathrm{maxsize}=25$, $1.0$--$2.1$\,h at $\mathrm{maxsize}=40$ and $1.9$--$3.7$\,h at $\mathrm{maxsize}=60$. At $\mathrm{maxsize}=100$, $400$ iterations did not finish within a four-hour limit in an earlier attempt, so those five fits were run at $250$ iterations and took $2.4$--$4.9$\,h. This is the practical reason the largest budget in that sweep is iteration-limited.

Debugging is also more difficult than for a fixed-architecture (e.g., simple PCA or polynomial fit) emulator. The discovered equations vary between runs unless the search is forced to be deterministic. Forcing determinism means setting \texttt{deterministic=True} with serial execution, giving up the four-way worker parallelism; we did not run the production budget in that configuration for wall-time speed reasons. These issues become more pronounced as the input dimension grows, which suggests that scaling PySR-based extraction to higher-dimensional emulators will likely require improved search heuristics or domain-specific factorization on the equation space.

\section{Conclusion}
\label{sec:conclusion}
In this work, we have presented symbolic regression models for the cosmological and astrophysical dependence of the one-dimensional Lyman-$\alpha$ power spectrum.  We trained our symbolic regression models on emulator spectra based on the multi-fidelity (across low- and high-resolution) PRIYA simulation suite. We used the multi-fidelity approach for PySR to learn equations that model the dependence of the power spectra on individual cosmological and astrophysical parameters. We evaluated performance in both high- and low-fidelity regimes. By building a framework of equations unique to each parameter, we are able to accurately model the PRIYA simulations in a multi-fidelity setting.

We have also shown that value accuracy is not sufficient for an emulator distilled into symbolic form to be used in a downstream analysis such as a Fisher forecast. An equation can reproduce the P1D flux power spectrum to the percent level and \textit{still} misrepresent its parameter derivative. We introduced a derivative-faithfulness diagnostic, ${\graderr}$, that scores each equation's parameter slope against the GP emulator. We also introduce a derivative-matching (Sobolev) training loss that repairs an unfaithful slope at the same search budget. Applied to the eleven PRIYA parameters at $z=3.6$, the Sobolev objective yields a derivative-faithful equation for nine of them, including the primordial amplitude and tilt, which the plain value loss leaves slope-biased. Only the Hubble parameter $h$ and the AGN feedback strength $\epsilon_{AGN}$, whose P1D imprints are weak or degenerate, resist.

We find that this gradient-matching repair comes from redefining the training objective rather than from a larger complexity search: value-loss is fragile to derivative-faithfulness in both the complexity budget and the random seed, while the Sobolev loss reaches faithfulness reliably at the smallest budget.
The pitfalls we document here in this work would be useful for similar symbolic-emulator work in the future.
Beyond accuracy, a key advantage of this approach lies in the potential interpretability of the learned equations. The resulting symbolic expressions make explicit how cosmological, astrophysical, and resolution parameters contribute to the power spectrum. In systematically varying single-dimensional relations, we can quantify the learned dependency of the emulator output on any given parameter, mitigating a ``black-box'' treatment of the emulator.  For example, the learned equation for $h$ contains $h$ only inside a single logarithm and is dominated by its dependence on resolution and wavenumber, so the slope it returns misses the emulator's altogether. In contrast, $d\tau_0$ enters its own equation linearly at leading order.

Our method provides a pathway towards assessing whether emulator corrections are meaningful in a physical sense rather than merely statistical. In particular, the learned dependence on simulation resolution suggests that symbolic regression can offer a new way to analyze how numerical effects propagate through Lyman-$\alpha$ forest 1D analysis.

Overall, this work shows that symbolic regression is a useful tool to complement cosmological emulators. By combining statistical emulators with analytical expressions, we are able to extract the underlying parameter dependence of a multi-fidelity model, begin interpretation of what is learned in a black-box emulator, and inform future computational efforts on high-need areas of parameter space.

\section*{Acknowledgments}
AB is supported by the University of Michigan's Summer Undergraduate Research Fellowship in the Department of Physics. MFH and CA are supported by the Leinweber Institute and DOE grant DE-SC0019193. S.B. and Y.Y. were supported in part by grant 63667 from the John Templeton Foundation. The opinions expressed in this publication are those of the
author(s) and do not necessarily reflect the views of the John Templeton Foundation. S.B. and Y.Y. were supported by NSF AST-2509639. 
GPT-5.6 Sol (OpenAI), Claude Opus 4.8, 5, 5.5 (Anthropic) were used to assist with software development, debugging, code review, and manuscript editing, including revisions to ensure that the presentation accurately reflected the underlying numerical analyses. The authors reviewed the AI-assisted code and text.  
\bibliographystyle{apsrev4-2}

\bibliography{oja_template}

\begin{appendix}

\section{PySR hyperparameter configuration}
\label{sec:apdx_kwargs}
The production fits use the operator and loss settings discussed in Section~\ref{sec:hyperparameters}: binary operators $\{+, -, \times, \mathrm{aq}\}$, unary operators $\{\exp, \log, x^2\}$, and the Sobolev loss with $\lambda=5$. The $x^y$ operator is nominally available but constrained out of the search. We use $\texttt{niterations}=200$, $\texttt{populations}=48$, and $\texttt{maxsize}=20$, with one model per parameter and redshift. The runs use four CPU cores in multithreaded mode and are therefore not fully deterministic. Full timing is given in Section~\ref{sec:cost}.

As a check on our equation selection rule, we also compare the knee against PySR's default \texttt{best} selection. Although the selected equations differ for eight of the eleven parameters at $z=3.6$, all eleven retain the same faithful or unfaithful classification under the Sobolev loss.

\section{Per-parameter equation table}
\label{sec:apdx_per1d_eqs}

Table~\ref{tab:per1d_eqs} lists the per-parameter equations from the production fits selected at the Pareto knee (Section~\ref{sec:hyperparameters}) under the Sobolev loss at $z=3.6$. The table also reports the equation complexity, training loss, and derivative error $\graderr$ defined in Eq.~\ref{eq:grad_err}. 

The knee equations generally have a higher training loss but lower complexity than the lowest-loss equations. The largest reductions in complexity are for $\alpha_q$, $h$, and $\Omega_M h^2$.

\begin{table*}[!t]
\caption{\label{tab:per1d_eqs} Per-parameter 1D PySR equations under the Sobolev loss at $z=3.6$. For each parameter, we report the Pareto-knee equation (Section~\ref{sec:hyperparameters}), its complexity, training loss, and derivative error $\graderr$ (Eq.~\ref{eq:grad_err}). We also report the lowest training loss from the full Pareto front before applying the validity filter, with its equation and complexity, for comparison. The equations are expressed in terms of the normalized inputs $\theta$ (parameter), $k$ (wavenumber), and $r$ (resolution).
}
\centering
\scriptsize
\newcommand{\eqcell}[1]{\parbox[t]{0.62\textwidth}{\ttfamily\raggedright #1\strut}}
\newcommand{\eqm}[1]{\parbox[t]{0.56\textwidth}{\raggedright\(#1\)\strut}}
\newcommand{\eqml}[1]{\eqm{#1}}
\begin{tabular}{c|>{}l|ccc|l}
\hline\hline
Param & selection & cmplx & loss & $\graderr$ & equation \\
\hline
$d\tau_0$ & knee & 19 & 0.0022 & 0.003 & \eqm{3.3965163\,\theta - \left(\frac{1.32706\,r}{\sqrt{1+(0.12813728\,\theta)^2}\,\sqrt{1+(k-0.08609456)^2}}\right)^{4} - 1.6309872} \\
 & lowest loss & {20} & {0.0021} & & \eqml{3.3990898\,\theta - \left(\frac{(1.2794386\,r)^{2}}{\left[1+(0.08799678\,\theta)^2\right]\sqrt{1+(k-0.075152986)^2}}\right)^{4} - 1.694589} \\
\hline
$\tau_0$ & knee & 20 & 0.0024 & 0.005 & \eqm{\frac{-0.5111359\,r}{\sqrt{1+k^2}} + 0.3953571 + \left(\theta - e^{-1.1378667\,\theta}\right)\left(2.010049 + 0.060667098\,r\right)} \\
 & lowest loss & & & & {same as knee} \\
\hline
${n_P}$ & knee & 18 & 9.75 & 0.160 & \eqm{\left(\frac{7.1688313}{\sqrt{1+\left(e^{\theta}\,(12.228655\,k)^2\right)^{4}}} - \theta - 2.8649871\right)(r - \theta)} \\
 & lowest loss & {20} & {8.90} & & \eqml{\left(\frac{7.173209}{\sqrt{1+\left(e^{\theta}\,(12.231928\,k)^2\right)^{4}}} - \theta - 2.8458169\right)(r - \theta) - r} \\
\hline
$A_P$ & knee & 19 & 13.55 & 0.155 & \eqm{\theta - \left(\frac{(1.3371198 - \theta\,e^{k})\,(r + 0.8623682)}{\sqrt{1+(e^{k^2} - 1.8447539)^2}}\right)^{2}} \\
 & lowest loss & {20} & {12.56} & & \eqml{\theta^{2} - \left(\frac{(1.2758272 - \theta\,e^{k})\,(r + 0.959206)}{\sqrt{1+(e^{k^2} - 1.8219881)^2}}\right)^{2}} \\
\hline
$z_{Hei}$ & knee & 18 & 0.72 & 0.084 & \eqm{\left(\frac{4.2322006\,\theta}{\sqrt{1+(1.8339454\,\theta r)^2}} - (r + 1.6339823)\right)\Big/\sqrt{1+\left(\frac{k^2}{1+\theta^2}\right)^{2}}} \\
 & lowest loss & {20} & {0.67} & & \eqml{\left(\frac{4.217855\,\theta}{\sqrt{1+(1.8287722\,\theta r)^2}} - e^{r}\right)\Big/\sqrt{1+\left(\frac{k^2}{1+\theta^4}\right)^{2}} - 0.37036908} \\
\hline
$z_{Hef}$ & knee & 20 & 18.88 & 0.045 & \eqm{\left(\frac{21.312387\,r}{\sqrt{1+(\theta+r)^2}} + 11.451199\,\theta\right)\left(\frac{\theta}{\sqrt{1+k^4}} - 1.7643517\right) + 13.6623745} \\
 & lowest loss & & & & {same as knee} \\
\hline
$\alpha_q$ & knee & 10 & 27.49 & 0.134 & \eqm{\theta\left(\frac{-68.50284}{\sqrt{1+(rk)^4}} + 65.33837\right)} \\
 & lowest loss & {20} & {9.56} & & \eqml{\left(\frac{1601.1101}{\sqrt{1+(rk)^{16}}} - 1598.3196\right)\frac{1.6343628 - \theta}{\sqrt{1+(r+\theta)^2}} - 3.5175257} \\
\hline
$h$ & knee & 10 & 422 & 1.000 & \eqm{\frac{-4.8548927}{\sqrt{1+\left(k + \log(\theta + r^2)\right)^{4}}}} \\
 & lowest loss & {19} & {386} & & \eqml{\left[1.8753233\left(k^2 + \log(\theta + r)\right)\right]^{2} - \theta^{2} + \frac{71.77047}{\sqrt{1+r^2}} - 66.73844} \\
\hline
$\Omega_M h^2$ & knee & 11 & 75.03 & 0.063 & \eqm{(k - 1.9741056\,\theta)^{2} - 11.100134\,r^{2}} \\
 & lowest loss & {20} & {68.46} & & \eqml{\left(\frac{r}{\sqrt{1+\theta^4}} - k^2 + 2.242719\,(\theta - 0.3432269)\right)^{2} - 17.277315\,r + 6.142204} \\
\hline
$z_{Hi}$ & knee & 19 & 8.58 & 0.167 & \eqm{\frac{42.282997 - 80.81628\,r}{\sqrt{1+\left(\frac{5.3947306}{\sqrt{1+\theta^2}}\right)^{2}}} - 3.0936897 + \left(\frac{\theta - 1.5468874}{\sqrt{1+r^2}}\right)^{4}} \\
 & lowest loss & {20} & {8.23} & & \eqml{\frac{33.885803 - 62.55704\,r}{\sqrt{1+\left[\exp\!\left(1.5835264/\sqrt{1+\theta^2}\right)\right]^{2}}} - 3.5045056 + \left(\frac{\theta - 1.5552967}{\sqrt{1+r^2}}\right)^{4}} \\
\hline
$\epsilon_{AGN}$ & knee & 20 & 12.28 & 0.771 & \eqm{\left(\log\!\left(\left(\frac{\theta}{\sqrt{1+k^2}} + r\right)^{2}\right)\right)^{2} - \frac{35.35155\,r^{4}}{\sqrt{1+\left(\frac{e^{k}}{\sqrt{1+\theta^2}}\right)^{2}}}} \\
 & lowest loss & & & & {same as knee} \\
\hline\hline
\end{tabular}
\end{table*}

\section{Derivative error threshold and the Sobolev loss}
\label{sec:apdx_sobolev}

Below we summarize the technical details how we calculate the derivative error. 
For each parameter $\theta_i$, we compare the derivatives of the PySR equation with respect to $\theta_i$, keeping all other parameters at their fiducial values (one parameter at a time). While the Sobolev loss compares these derivatives across the training samples, the derivative error $\graderr$ (Eq.~\ref{eq:grad_err}) evaluates their agreement only at $\theta_i=\theta_{i,\mathrm{fid}}$, across the $k$-bins. We use the same central finite-difference stencil for both PySR and GP. The score is evaluated only where the GP slope is non-negligible, with $|\partial_\theta \log P^{\mathrm{GP}}| \ge 10^{-3}$ of the peak response for that parameter. This avoids taking a relative error against a slope that is effectively zero. In practice, the cut removes almost nothing. All reported scores retain all $48$ $k$-bins, except for $\epsilon_{AGN}$ at $z=2.6$, where $47$ are retained.

We calculate the derivative error for each $k$-bin and take the median across the retained bins as $\graderr$. This measures how well the equation reproduces the GP parameter gradient for a typical $k$-bin, without being strongly affected by a few bins with large errors. 

We classify an equation as \emph{faithful} if $\graderr<0.25$. Under the Sobolev loss, this threshold clearly separates the equations at $z=3.6$: across five random seeds, the nine faithful parameters have errors no larger than $0.246$, while the other two, $h$ and $\epsilon_{\rm AGN}$, have errors above $0.59$. The improvement is also clear at $z=4.2$, but less so at $z=2.6$, where the spectral tilt $n_P$ has an error of $0.255$, just above the threshold. The classification is also more sensitive to the threshold under the value-only loss. We therefore report the measured $\graderr$ values alongside the classifications for transparency.


We implement the Sobolev loss in PySR by comparing both the predicted values and parameter gradients of the symbolic equations with those of the GP emulator at each resolution. The loss is evaluated in the normalized log-$P$ training space with $\lambda=5$. The implementation is provided in \texttt{src/priya\_forecast/sobolev\_loss.py}.

All per-candidate scores are saved in the files \texttt{grad\_faith\_<param>.csv} alongside each Pareto front, generated by \texttt{scripts/eval\_grad\_faithfulness.py}. The diagnostic figures are produced by \texttt{scripts/make\_diagnostic\_figs.py}.

\section{{Multi-parameter validation errors}}
\label{sec:apdx_multid}
\begin{table}[H]   
\caption{
Multi-parameter additive-Taylor combine versus GP at $z=3.6$ ($256$ Sobol points per combination): mean and maximum relative error over ($\theta$, $k$), and the $k$ of the worst bin. Selection: the lowest-loss equation among the Fisher-safe candidates (Section~\ref{sec:hyperparameters}); the derivative-faithfulness threshold is applied on top of it but does not change the selection for any parameter shown here.
}
\label{tab:multid}
\begin{ruledtabular}
\centering
\begin{tabular}{c|c|ccc}
Dim & Parameters & mean (\%) & max (\%) & $ {k_{\rm worst}}$ [s/km] \\
\hline
2D & $d\tau_0$--$A_P$ & 0.73 & 2.84 & 0.0035 \\
2D & $z_{Hei}$--$\alpha_q$ & 0.68 & 8.70 & 0.0010 \\
2D & $A_P$--${n_P}$ & 0.88 & 4.83 & 0.0010 \\
2D & $\tau_0$--${n_P}$ & 0.34 & 2.94 & 0.0010 \\
2D & $z_{Hei}$--$z_{Hef}$ & 0.39 & 4.47 & 0.0022 \\
2D & $\tau_0$--$A_P$ & 0.60 & 3.45 & 0.0035 \\
3D & $d\tau_0$--$A_P$--${n_P}$ & 0.86 & 4.88 & 0.0010 \\
3D & $z_{Hei}$--$z_{Hef}$--$\alpha_q$ & 0.80 & 9.79 & 0.0010 \\
3D & $\tau_0$--$A_P$--${n_P}$ & 0.79 & 4.95 & 0.0010 \\
3D & $A_P$--${n_P}$--${\Omega_M h^2}$ & 0.92 & 5.50 & 0.0013 \\
\end{tabular}
\end{ruledtabular}
\end{table}

\end{appendix}

\end{document}